\documentclass[aps,prb,preprint,floatfix,superscriptaddress]{revtex4-2}

\usepackage{graphicx}
\usepackage{amsmath,amssymb,amsfonts}
\usepackage{url,hyperref}
\usepackage{multirow}
\usepackage[version=4]{mhchem}
\usepackage{subcaption}
\usepackage{mathtools}
\usepackage{verbatim}
\usepackage{titlesec}
\usepackage{titlecaps}
\usepackage{enumitem}
\usepackage{color}
\usepackage[T1]{fontenc}
\usepackage[utf8]{inputenc}

\usepackage{xcolor}
\usepackage{soul}
\sethlcolor{white}

\graphicspath{{Figures/}}

\usepackage{chngcntr}  
\usepackage{caption}

\newcounter{SIfigure}
\renewcommand{\theSIfigure}{S\arabic{SIfigure}}
\newcounter{SItable}
\renewcommand{\theSItable}{S\arabic{SItable}}
\setcitestyle{super,comma,sort&compress}       
\begin{document}
\title{Phase Stability and Transformations in Lead Mixed Halide Perovskites from Machine Learning Force Fields}

\author{Xia Liang}
\affiliation{Department of Materials, Imperial College London, South Kensington Campus, London SW7 2AZ, UK\\}

\author{Johan Klarbring}
\affiliation{Department of Physics, Chemistry and Biology (IFM), Link\"{o}ping University, SE-581 83, Link\"{o}ping, Sweden}

\author{Aron Walsh}
\email{a.walsh@imperial.ac.uk}
\affiliation{Department of Materials, Imperial College London, South Kensington Campus, London SW7 2AZ, UK\\}

\date{\today}

\vspace*{4em}

\begin{abstract}
Lead halide perovskites (\ce{APbX3}) offer tunable optoelectronic properties but feature an intricate phase stability landscape. Here, we employ on-the-fly data collection and an equivariant message-passing neural network potential to perform large-scale molecular dynamics of three prototypical perovskite systems: \ce{CsPbX3}, \ce{MAPbX3}, and \ce{FAPbX3}. Integrating these simulations with the \textsc{PDynA} analysis toolkit, we resolve both equilibrium phase diagrams and the dynamic structural evolution under varying temperatures and halide mixtures. Our findings reveal that A-site cations strongly modulate tilt modes and phase pathways: \ce{MA+} effectively “forbids” the $\beta$-to-$\gamma$ transition in \ce{MAPbX3} by requiring extensive molecular rearrangements and crystal rotation, whereas the debated low-temperature phase in \ce{FAPbX3} is predicted to be best represented as an $Im\bar{3}$ ($a^{+}a^{+}a^{+}$) cubic phase. Additionally, small changes in halide composition and arrangement, from uniform mixing to partial segregation, alter octahedral tilt correlations. Segregated domains can even foster anomalous tilting modes that impede uniform phase transformations. These results highlight the multi-scale interplay between the cation environment and halide distribution, offering a rational basis for tuning perovskite architectures toward improved phase stability. 

\end{abstract}

\maketitle
\pagebreak

\section{Introduction}
Lead halide perovskites have the general formula \ce{APbX3} (where \ce{A} is a monovalent cation and \ce{X} is a halide). They have emerged as some of the most transformative and versatile materials in modern optoelectronics since their first fabrication half a century ago.~\cite{early_cspbi3_moller,early_mapi_weber} They have demonstrated utility in light-emitting diodes, sensors, lasers, and especially photovoltaics,~\cite{perovskite_LEDs_lin,MA_lasing_xing,perovskite_sensors_xie,perov_sc_first2009} where power conversion efficiencies have risen from about 4\% to over 25\% in just a decade.~\cite{perov_sc_first2009,compositional_engineering_nature_jeon,perov_tandom_25.2_sahli} Their strong optical absorption, tunable band gaps, and high carrier mobilities make them highly attractive for device integration, while low-cost solution-based fabrication promises scalable production routes.~\cite{aron_origin_performance,mapi_defects_yan,cspbbr3_absorption_maes,perov_cation_Eg_tuning_amat} Nonetheless, these advantages come with challenges: moisture sensitivity, trap-state formation, thermal/photo-instability, and lead toxicity all necessitate further compositional engineering.~\cite{mapi_instability_2015_conings,hhp_instability_park,electronics_segregation_makarov,giwaxs_segregation_merten,exp_trap1} This duality—high performance with inherent drawbacks—has fueled extensive research into both fundamental mechanisms and advanced optimization strategies.

Of the many possible monovalent cations for the A-site, caesium (Cs), formamidinium (FA, \ce{CH(NH2)2+}), and methylammonium (MA, \ce{CH3NH3+}) are most frequently used because their ionic radii ensure a suitable tolerance factor for stable, three-dimensional octahedral frameworks.~\cite{disorder_black_cspbi3_arthur,mapbbr_sim2016even,azr_perovskite_mantas,FA_based_halide_li_2018} Other cations, such as ethylammonium and methylhydrazinium, have also been investigated but remain less common.~\cite{EAPbI3_exp_zhang,MHyPbX3_exp_sieradzki} Specifically, Cs/FA/MA lead iodides display band gaps between 1.51 and 1.72\,eV, whereas their bromide counterparts range from 2.24 to 2.40\,eV,~\cite{electronics_tin_lead_perovskites_tao,three_cations_bromide_bandgap_giovanni,mapbibr3_exp_krict_2013,fapi_exp_science_yang} an interval well suited for single-junction or tandem solar cells. Both Cs- and MA-based perovskites exhibit a well-documented three-phase sequence (cubic, tetragonal, and orthorhombic) as temperature changes,~\cite{ma-fa_phase_diagram_adrian,mixed_lead_halide_perovskite_review_Mantas,disorder_black_cspbi3_arthur,abinit_fapi_mapi_scirep_kim} whereas the phase stability of FA-based compounds remains actively debated.~\cite{black_stabilization_fapi_cspbi3_sofia,structure_optical_fapi_douglas,charge_carrier_lifetime_hybrid_chen} Beyond static phases, all these perovskites display dynamic local distortions that can shape optoelectronic properties.~\cite{Cs_local_structure_baldwin,dynamic_domian_milos} For example, FA-based systems can exhibit dynamic tilt modulations that favour hexagonal phases,~\cite{perov_tilting_stable_science} whereas transient ferro- or antiferroelectric domains in MA-based perovskites enhance carrier lifetimes and influence halide migration.~\cite{MA_local_order_2023_toney}

Alongside tuning the A-site, mixing different halides (typically \ce{I} and \ce{Br}) at the X-site provides an additional route for tailoring band gaps and absorption edges, but it also raises the risk of halide segregation.~\cite{doubly_science_mcmeekin,MAPbICl_mixing_carrier_transport_chen,mixed_lead_halide_perovskite_review_Mantas,Cs_mix_halide_protesescu} 
Such segregation can degrade charge transport, compromise operational stability, and reduce overall device performance.~\cite{exp_trap1,Cs_segregation_degrade_huang,uv_degredation_mechanism_lang} While the halides in bulk structure without external impact remain evenly distributed~\cite{pb207_nmr_karmakar,halide_mixing_MA_fykouras}, illumination, elevated temperatures, or humidity can promote segregation over seconds to minutes, far exceeding the typical nanosecond timescales of standard molecular dynamics (MD) simulations.~\cite{steric_suppress_grater,illumination_segregate_yun,Cs_segregation_degrade_huang,nanoribbon_segregate_akash,kMC_segregation_kuno} As a result, local band-gap inhomogeneities emerge, often manifesting as a red shift in absorption that undermines device performance.~\cite{perov_tilting_stable_science,electronics_segregation_makarov} Mitigation strategies include chemical additives, targeted light excitation, or self-healing processes,~\cite{Cs_mix_halide_protesescu,laser_remixing_doubly_okrepka,ligand_suppress_segregattion_ghorai} as well as compositional engineering.~\cite{MA_add_Cs_stability_beal,perov_tilting_stable_science} Modeling mixed sites in crystals requires careful sampling of the configurational space, often achieved via approaches such as enumeration over unique configurations (e.g. site-occupancy disorder~\cite{configurational_modelling_deleeuw}) or construction of a representative single configuration, special quasirandom structures (SQS)~\cite{sqs_original}.

Conventional first-principles techniques based on density functional theory (DFT) have provided valuable insights into perovskite structures and properties,~\cite{atomistic_models_perovskite_walsh,hhp_perspective_whalley,hhp_overview_jarvist} but are generally limited by computational expense, restricting the size and timescale of simulations. These constraints pose a substantial challenge for exploring long-range tilt correlations, large supercells, and segregation phenomena in mixed-halide perovskites. Machine-learning (ML) methods have gained traction for their ability to navigate broad compositional spaces, accelerate structure generation, and offer near \emph{ab initio} accuracy in atomic simulations at dramatically reduced cost.~\cite{material_ml_overview_aron,mattergen_original,ml_screening_review_froudakis,ml_discovery_from_failed_norquist} Recent developments in machine learning force fields (MLFFs) achieve near \textit{ab initio} accuracy at greatly reduced computational cost, enabling the study of phase transitions and defect-driven processes in large supercells.~\cite{mlp_review_jorg_2016,mlp_review_csyani_2021,vaspmlff_original,m3gnet_original,allegro_original,mace_original,mattersim_original} We adopt the MACE architecture, building on the atomic cluster expansion (ACE) framework~\cite{ace2019drautz,ace2022dusson}. By employing higher-order equivariant message passing and a deep-learning optimizer, MACE effectively captures many-body atomic interactions and is thus well suited for complex local tilt behaviours and long-range structural ordering in mixed-halide perovskites.~\cite{mace_original,mace_mp0}

In this paper, we detail our on-the-fly DFT data generation and MACE training strategy, then show how these potentials enable large-scale MD simulations for homogeneous halide mixing and artificial segregation. We present the calculated phase diagrams and investigate temperature-dependent phase transformations, emphasizing the roles of A-site cations, halide composition, and dynamic domain formation. Finally, we discuss the broader implications of these findings for perovskite engineering and highlight how advanced ML-driven simulations and structural metrics can inform the design of stable, high-performance materials.

\section{Methods} 

We combine two force fields in this study: one for training data generation and another for large-scale production runs. The first MLFF, based on Gaussian process regression~\cite{gaussian_process_volker}, is integrated within the Vienna \textit{Ab initio} Simulation Package (\textsc{VASP})~\cite{vasp_original1,vasp_original2}. It facilitates on-the-fly DFT data collection, selecting local configurations using Bayesian error estimation during MD simulations. 
The second MLFF employs an equivariant deep learning interatomic potential~\cite{mace_original}, trained on the dataset collected in the first step. All structural analyses, such as trajectory evaluation and training dataset inspection, are carried out with the \textsc{PDynA} package~\cite{pdyna2023liang}.

\subsection{Training Strategy and DFT Data Collection}

We construct the structural training dataset in two stages. First, we prepare initial configurations and select simulation temperatures to comprehensively sample both chemical composition and temperature ranges for each mixed halide system (\ce{CsPbX3}, \ce{MAPbX3}, and \ce{FAPbX3}). For each system, we generate three initial structures from known polymorphs $\alpha$, $\beta$, and $\gamma$, as summarized in Table~\ref{profiles}. We then account for halide composition by substituting 5, 9, 15, and 19 iodine atoms with bromine in a $2\times2\times2$ supercell using special quasi-random structures from the ICET package~\cite{icet_2019}. Each system therefore undergoes 20 on-the-fly learning MD simulations, ensuring broad coverage of structural and compositional variations.

\begin{table}[ht]
\centering
\caption{Training profiles including crystal phase (and corresponding tilting mode) of the initial structure and the MD temperature. }
\begin{tabular}[t]{c@{\hskip 0.3in}c@{\hskip 0.2in}c@{\hskip 0.2in}c@{\hskip 0.2in}c@{\hskip 0.2in}c}
\multirow{2}{*}{Material} & \multicolumn{5}{c}{Temperature in Kelvin (Phase)} \\
\cline{2-6}
 & 1 & 2 & 3 & 4 & 5 \\
\hline\hline
\textbf{\ce{CsPbX3}} & 700 ($ \alpha $) & 600 ($ \alpha $) & 510 ($ \beta^{+} $) & 325 ($ \beta^{+} $) & 150 ($ \gamma^{-} $) \\
\textbf{\ce{MAPbX3}} & 450 ($ \alpha $) & 350 ($ \alpha $) & 200 ($ \beta^{-} $) & 150 ($ \gamma^{-} $) & 100 ($ \gamma^{-} $) \\
\textbf{\ce{FAPbX3}} & 450 ($ \alpha $) & 320 ($ \alpha $) & 150 ($ \beta^{+} $) & 100 ($ \gamma^{+} $) & 50 ($ \gamma^{+} $) \\
\hline\hline
\textbf{Phase} & $ \alpha $ & $ \beta^{-} $ & $ \beta^{+} $ & $ \gamma^{-} $ & $ \gamma^{+} $ \\
\textbf{Tilting Mode} & $ a^{0}a^{0}a^{0} $ & $ a^{0}a^{0}c^{-} $ & $ a^{0}a^{0}c^{+} $ & $ a^{-}a^{-}c^{+} $ & $ a^{+}a^{+}a^{+} $ \\
\hline
\end{tabular}
\label{profiles}
\end{table}%

On-the-fly learning proceeds via Bayesian error estimation~\cite{vaspmlff_original}, which requests DFT calculations only when predicted errors exceed a threshold. Each DFT evaluation records energy, forces, and stress tensors to refine the force field. MD simulations use the isothermal-isobaric (NpT) ensemble at 1\,bar on $2\times2\times2$ supercells. For \ce{CsPbX3}, this corresponds to 40 atoms; for \ce{MAPbX3} and \ce{FAPbX3}, 96 atoms. We apply Langevin thermostats with atomic and lattice friction constants of 10\,$\mathrm{ps}^{-1}$. Timesteps are set to 1.0\,$\mathrm{fs}$ for inorganic systems and 0.5\,$\mathrm{fs}$ for organic systems, each simulation runs for 100{,}000 steps.

Across the 20 simulations, we obtain roughly 18{,}000, 11{,}200, and 7{,}080 DFT snapshots for \ce{CsPbX3}, \ce{MAPbX3}, and \ce{FAPbX3}, respectively. A set of DFT single-point calculations are performed on these collected structures to eventually construct the training set for the force field. All DFT calculations employ projector-augmented wave (PAW) potentials, the r$^2$SCAN exchange-correlation functional~\cite{r2scan_original1}, an electronic convergence threshold of $10^{-5}\,\mathrm{eV}$, and a plane-wave cutoff of 550\,$\mathrm{eV}$. A $2\times2\times2$ $\Gamma$-centered $k$-point grid is used for training, consistent with previous work~\cite{pdyna2023liang}.

\subsection{MACE and Molecular Dynamics}
We then adopt the MACE architecture, an equivariant deep learning framework based on the ACE structural descritor~\cite{ace2019drautz,mace_original}. MACE leverages higher-order equivariant descriptors and a deep-learning optimizer to model many-body atomic interactions. Separate MACE models are constructed for \ce{CsPbX3}, \ce{MAPbX3}, and \ce{FAPbX3}, each using a radial cutoff of 7\,\AA, two message passing layers (64 channels), a message equivariance $L=0$, a message body order of 4, and a spherical harmonics embedding of order 3. The data are split into $80\%$ for training, $10\%$ for testing, and $10\%$ for validation, and shuffled before each run. We minimize a $1:2:2$ weighted loss function for energy, forces, and stress, using the Adam optimizer~\cite{pytorch_2019} with a batch size of 5 and a learning rate of 0.002. Training completes after 1{,}000, 1{,}200, and 1{,}200 epochs for \ce{CsPbX3}, \ce{MAPbX3}, and \ce{FAPbX3}, respectively. The energy weight in the loss function was increased after 800 epochs, once forces are sufficiently optimised. RMSE values obtained for unseen validation data are provided in the Supplementary Information (Table \ref{errors}). 

For the MD simulations, relaxed pristine structures are scaled to the target size. In organic systems, each A-site molecule is randomly rotated to cover a broad range of orientations. Halide mixing follows two schemes: Firstly, \textit{homogeneous}, where bromine and iodine atoms are randomly assigned to X-sites. Fig. \ref{si:system_size} shows that as the supercell size increases, the difference between SQS-based and random halide placements becomes negligible. Secondly, \textit{heterogeneous}, where a custom Monte Carlo algorithm promotes halide segregation. In the heterogeneous scheme, we randomly select an I--Br pair at each iteration and calculate the surrounding halide concentrations before and after a hypothetical swap. The swap is accepted only if it increases local halide concentration. We quantify the degree of overall segregation via the segregation parameter $k_{seg}$,
\begin{equation}
\label{eq:segregate}
 k_{seg} = \frac{2 N_{0} + N_{1} + N_{5} + 2 N_{6}}{2N_{\text{total}}}
\end{equation}
where $N_n$ is the number of octahedra with $n$ bromine atoms, and $N_{\text{total}}$ is the total octahedra count. This parameter provides a robust metric for analyzing halide distribution across structures. The relationship between parameter $k_{seg}$ and the segregated structure is shown in Fig. \ref{si:segregate}.

We perform the large-scale simulations in LAMMPS~\cite{lammps_original} using a $14\times14\times14$ pseudo-cubic cell (13{,}720 atoms) for inorganic systems and a $10\times10\times10$ cell (12{,}000 atoms) for organic systems unless stated otherwise. The MD timestep is $1\,\mathrm{fs}$. Runs are executed on an NVIDIA A100 GPU with production rates of approximately 0.16, 0.19, and 0.12\,ns/day for \ce{CsPbX3}, \ce{MAPbX3}, and \ce{FAPbX3}, respectively. Each system equilibrates in the NpT ensemble at the target temperature for 300\,ps, with data collected over the final 200\,ps. We employ a Nosé-Hoover thermostat and barostat, with damping parameters of $100\,\mathrm{fs}$ and $200\,\mathrm{fs}$, respectively. For organic perovskites, the slower rotational dynamics of A-site molecules necessitate three starting configurations related to the molecular orientations. Heating and cooling runs begin from the final snapshot of an equilibration at the initial temperature. To ensure statistical significance, independent runs were performed with different thermalization seeds and equilibration times. The temperature gradient of all heating and cooling runs is set to 0.2 K/ps.

\subsection{Perovskite Dynamics Analysis}

We employ the \textsc{PDynA} package~\cite{pdyna2023liang} for structural dynamics analysis of large-scale perovskite MD trajectories. 
\textsc{PDynA} extracts tilt angles, distortion modes, molecular orientations, pseudocubic lattice parameters, and tilt correlations from each snapshot.
\textsc{PDynA} has been applied to a diverse array of perovskite compositions and phases~\cite{Cs_local_structure_baldwin,dynamic_domian_milos,picosecond_lifetime_MA_Eduardo,azr_perovskite_mantas,pdyna2023liang}, offering insights into how local structural fluctuations shape the macroscopic properties. Here, we will make use of two key structural descriptors: tilting correlation polarity (TCP) and tilting correlation length.

The TCP along a specific direction $\alpha$ takes values from $-1$ to $1$, indicating the nature of the $\alpha$-tilt correlation along $\alpha$ direction, analogous to the Glazer notation superscripts.~\cite{glazer_original} In particular, $-1$, $0$, and $+1$ correspond to the superscripts $-$, $0$, and $+$, respectively, while non-integer values reflect imperfect correlations—typical of dynamical systems. To begin, we compute a correlation pair $r_{\alpha,\beta}^{(k)}$ (where $\alpha$ and $\beta$ are any of the $x$, $y$, or $z$ principal axes) as the normalized product of the $\alpha$-tilt $\theta_\alpha$ of one octahedron and that of another octahedron $k$ unit cells away in the $x$ direction:
\begin{gather}
 \mathclap{r_{\alpha,x}^{(k)}(t;\mathbf{n}) =  \frac{\theta_\alpha(t;n_x,n_y,n_z)\,\theta_\alpha(t;n_x+k,n_y,n_z)}{\sqrt{\bigl|\theta_\alpha(t;n_x,n_y,n_z)\,\theta_\alpha(t;n_x+k,n_y,n_z)\bigr|}} } \label{eq:tcp1} 
 \\
 \delta_{\alpha} = \frac{n_{\alpha}^{+} - n_{\alpha}^{-}}{n_{\alpha}^{+} + n_{\alpha}^{-}} \label{eq:tcp2} 
\end{gather}
Here, $\mathbf{n} = (n_x,n_y,n_z)$ denotes the integer coordinates of an octahedron within the 3D supercell, and $t$ indicates a fixed time step in the MD trajectory. The quantity $\delta_{\alpha}$ then compares the counts of positive ($r_{\alpha,\alpha}^{(1)}>0$) versus negative ($r_{\alpha,\alpha}^{(1)}<0$) correlation pairs for first-nearest neighbours in the normal direction (where $\alpha=\beta$), yielding $n_{\alpha}^+$ and $n_{\alpha}^-$. 

Next, we compute the spatial correlation function of tilting, $R_{\alpha,\beta}(k)$, as an average over all relevant octahedron pairs in the supercell and over time frames ($\langle\cdots\rangle_{\mathbf{n},t,\pm k}$):
\begin{gather}
  \mathclap{R_{\alpha,x}(k) = C\,\Bigl\langle \bigl|\theta_\alpha(t;n_x,n_y,n_z)\,\theta_\alpha(t;n_x+k,n_y,n_z)\bigr| \Bigr\rangle_{\mathbf{n},t,\pm k}} \label{eq:corr_length1}
  \\
  R_{\alpha,\beta}(k) = \exp\Bigl(-\frac{k}{\xi_{\alpha,\beta}}\Bigr) \label{eq:corr_length2}
\end{gather}
where $C$ is a normalization constant enforcing $R_{\alpha,\beta}(0) = 1$. By fitting $R_{\alpha,\beta}(k)$ to the exponential decay in Eq.~\ref{eq:corr_length2}, we obtain a $3\times3$ correlation length tensor $\boldsymbol{\xi}$. Each element $\xi_{\alpha,\beta}$ represents the fitted correlation length of the $\alpha$-tilt along the $\beta$ direction in the structure. Together, TCP and correlation length provide a rigorous framework for identifying phase boundaries, detecting tilt domain nucleation, and tracking local structural variations under varying conditions of composition and temperature.

\section{Results and Discussion}

The structural behaviours of \ce{CsPbX3}, \ce{MAPbX3}, and \ce{FAPbX3} are explored in a range of initial configurations. Given the importance of halide arrangement, we compare two positioning schemes: \emph{homogeneous} and \emph{heterogeneous}. In the \emph{homogeneous} scheme, halides are randomly distributed across X-sites based on their overall concentration. The \emph{heterogeneous} mixing scheme involves the simulation of structures with locally segregated halides.

\subsection{Unmixed Systems}
\begin{figure*}[htb] 
    \centering
    \includegraphics[width=0.7\textwidth]{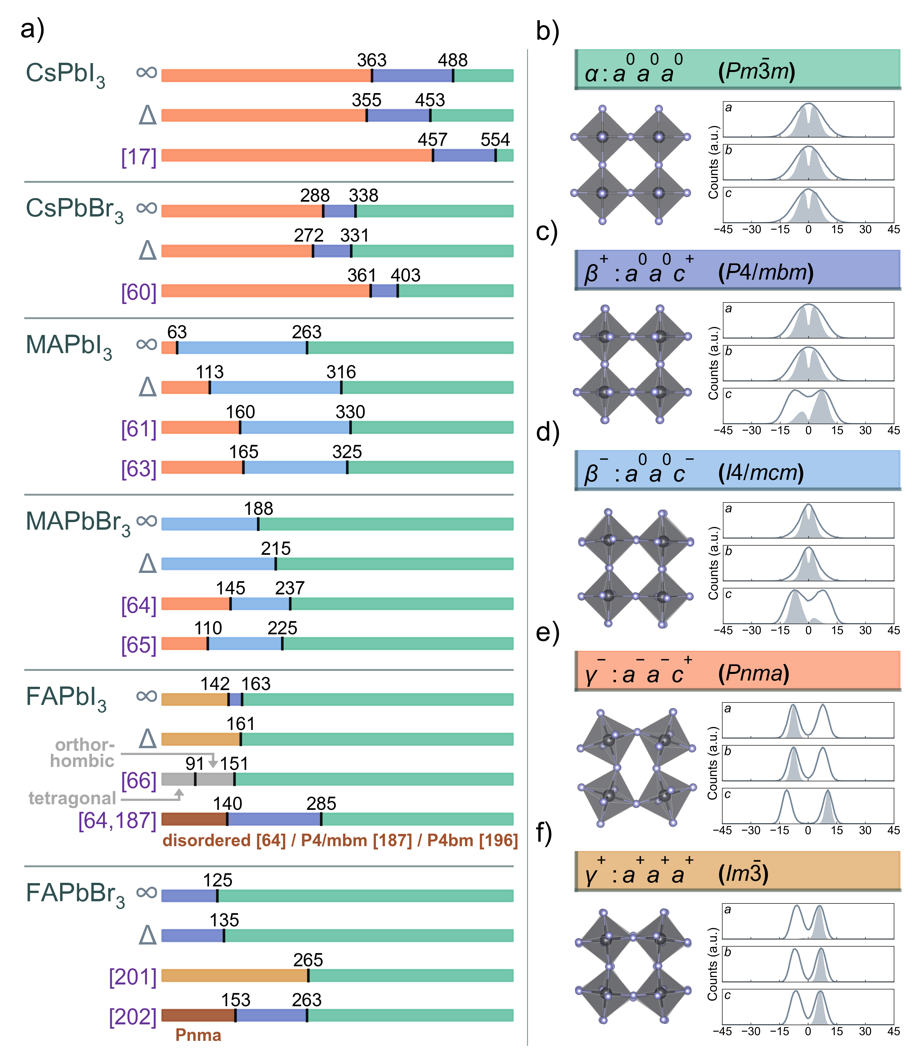}
    \caption{(a) Predicted crystal phases and transition temperatures ($\infty$ and $\Delta$ indicate transition temperatures obtained from static and transient calculations, respectively) of \ce{CsPbI3}, \ce{CsPbBr3}, \ce{MAPbI3}, \ce{MAPbBr3}, \ce{FAPbI3} and \ce{FAPbBr3}, a range of experimental references~\cite{disorder_black_cspbi3_arthur,phase_trans_ref_cspbbr3_1,abinit_fapi_mapi_scirep_kim,ma-fa_phase_diagram_adrian,mixed_lead_halide_perovskite_review_Mantas,pdyna2023liang,black_stabilization_fapi_cspbi3_sofia,structure_optical_fapi_douglas,charge_carrier_lifetime_hybrid_chen,dynamic_domian_milos,fa_pb-sn_i3_crystal_Emily} are included for comparison. (b--f) Tilt patterns of five unique phases. Each panel includes a tilt-correlation diagram containing three principal axes, where solid lines show tilt angle distributions and shaded areas indicate nearest-neighbour tilt correlations, highlighting in-phase or out-of-phase correlations. }
    \label{res:endpoints}
\end{figure*}

We begin by validating our workflow on pure (unmixed) \ce{CsPbX3}, \ce{MAPbX3}, and \ce{FAPbX3}. Fig. \ref{res:endpoints}a summarizes stable phases and approximate phase transition temperatures. Five distinct tilting modes (or symmetries) appear (see Fig. \ref{res:endpoints}b--f). For \ce{CsPbX3}, three phases emerge: cubic ($\alpha$, $a^{0}a^{0}a^{0}$), tetragonal ($\beta^{+}$, $a^{0}a^{0}c^{+}$), and orthorhombic ($\gamma^{-}$, $a^{-}a^{-}c^{+}$). These phases were consistent with training steps and prior studies.~\cite{Cs_mix_halide_protesescu,disorder_black_cspbi3_arthur,phase_trans_ref_cspbbr3_1} Phases were classified using three-dimensional octahedral tilting and tilt correlation along principal axes, analogous to Glazer notations~\cite{glazer_original}. The superscripts indicate whether neighbouring octahedra tilt \emph{in-phase} or \emph{out-of-phase}, but instead of a single plus or minus sign we use correlation magnitude and type (the tilting correlation polarity or TCP values). The dynamic effects near phase boundaries produce imperfect correlations, highlighting the need for this extended notation. Similar transitions occur in \ce{CsPbBr3}, though at slightly lower temperatures.

\begin{figure*}[htb]
    \centering
    \refstepcounter{SIfigure}
    \includegraphics[width=0.95\textwidth]{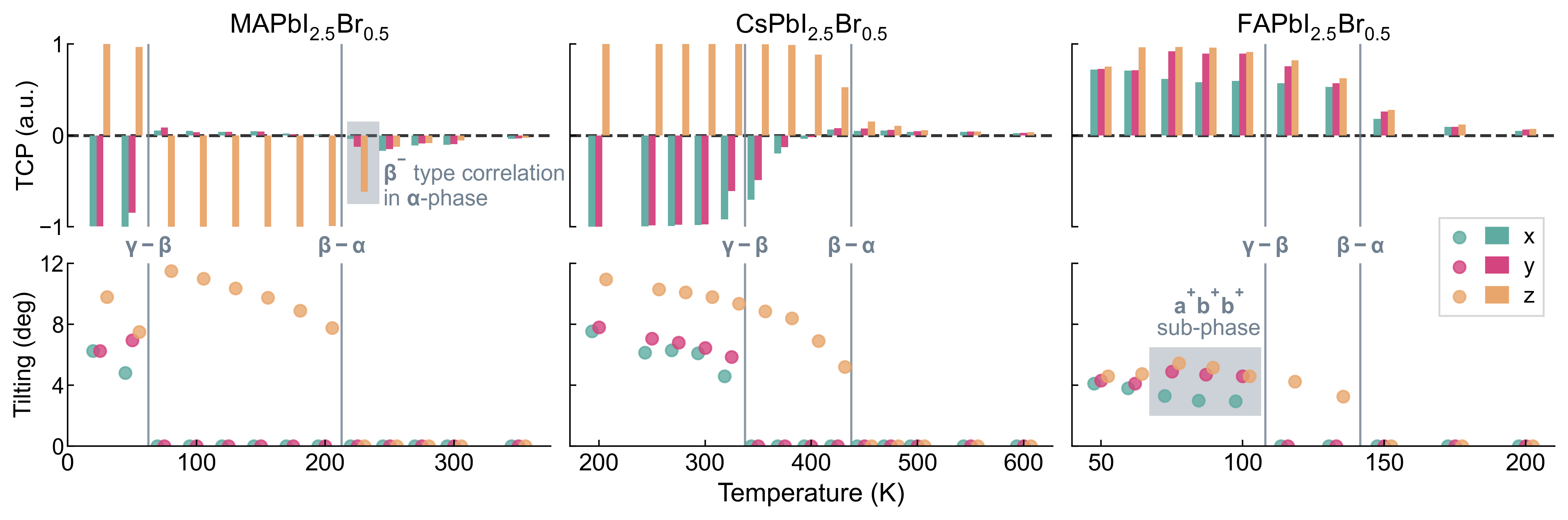}
    \caption*{Fig. \theSIfigure: TCP values and tilt angles on the three principal axes of \ce{MAPbI_{2.5}Br_{0.5}}, \ce{CsPbI_{2.5}Br_{0.5}}, and \ce{FAPbI_{2.5}Br_{0.5}} at various temperatures. The corresponding phase transition temperatures are indicated with vertical lines. } 
    \label{si:profiles}
\end{figure*}

\ce{MAPbX3} exhibits three distinct crystallographic phases: $\alpha$ ($a^{0}a^{0}a^{0}$), $\beta^{-}$ ($a^{0}a^{0}c^{-}$), and $\gamma^{-}$ ($a^{-}a^{-}c^{+}$). These assignments closely match prior experimental and computational work, showing reasonable agreement with known transition temperatures.~\cite{mapbbr_sim2016even,abinit_fapi_mapi_scirep_kim,mixed_lead_halide_perovskite_review_Mantas} Notably, in the iodide-rich end, the orthorhombic $\gamma$-phase emerges only at very low temperatures, whereas it remains unstable in bromide-rich compositions. Within the temperature range of 180 to 220~K, the structure of \ce{MAPbBr3} remains in the $\alpha$-phase; however, the TCP value of one axis becomes negative (Fig. \ref{si:profiles}), suggesting the formation of a preferred orientation for the crystallization of the $\beta$-phase. This observation aligns with our previous findings on \ce{MAPbBr3} with a different force field.~\cite{pdyna2023liang}

The larger and more anistropic \ce{FA} cation generally exhibits more complex phase behavior, which can become frustrated at lower temperatures. Indeed, there are multiple reported low-temperature structures, including disordered tilts, $P4/mbm$, $P4bm$, etc.~\cite{mixed_lead_halide_perovskite_review_Mantas,black_stabilization_fapi_cspbi3_sofia,structure_optical_fapi_douglas,charge_carrier_lifetime_hybrid_chen}. Here, \ce{FAPbI3} displays cubic phases ($\alpha$, $a^{0}a^{0}a^{0}$ at high temperature and $\gamma^{+}$, $a^{+}a^{+}a^{+}$ at low temperature) with a narrow tetragonal intermediate ($\beta$, $a^{0}a^{0}c^{+}$). In addition, a subtle sub-phase with tilting mode $a^{+}b^{+}b^{+}$ is found within the temperature range of 70~K to 100~K (Fig. \ref{si:profiles}). However, due to the insignificance of the difference between this sub-phase to the low-temperature $\gamma^{+}$ phase, it is identified as the latter. A similar sequence appears in \ce{FAPbBr3}, though its $\gamma$-phase does not form under these conditions. This particular sequence of phases aligns with a recent experimental sample characterised by X-ray diffuse scattering and inelastic neutron spectroscopy, in which a $Im\bar{3}$ symmetry with an $a^{+}a^{+}a^{+}$ tilting mode is identified.~\cite{dynamic_domian_milos}
As a caveat, we do not exhaustively explore the low-temperature landscape, which would require appropriate sampling techniques to overcome the sluggish dynamics of these systems. 

Although there is a general trend that the predicted transitions are shifted to lower temperatures relative to experiments. These differences are very small in terms of an energy scale (several meV/atom). There is a known softening effect of universal machine learning potentials~\cite{universal_softening_deng,softening_finetune_deng}, which may also be present for our potentials that are trained to describe a large compositional space. Importantly, the relative phase stabilities and symmetry-breaking behaviours remain well-described and physically meaningful. Our focus is on relative phase relationships rather than absolute temperatures, preserving the overall trends in octahedral tilting modes and cation dynamics.

\begin{figure*}[htb]
    \centering
    \includegraphics[width=0.9\textwidth]{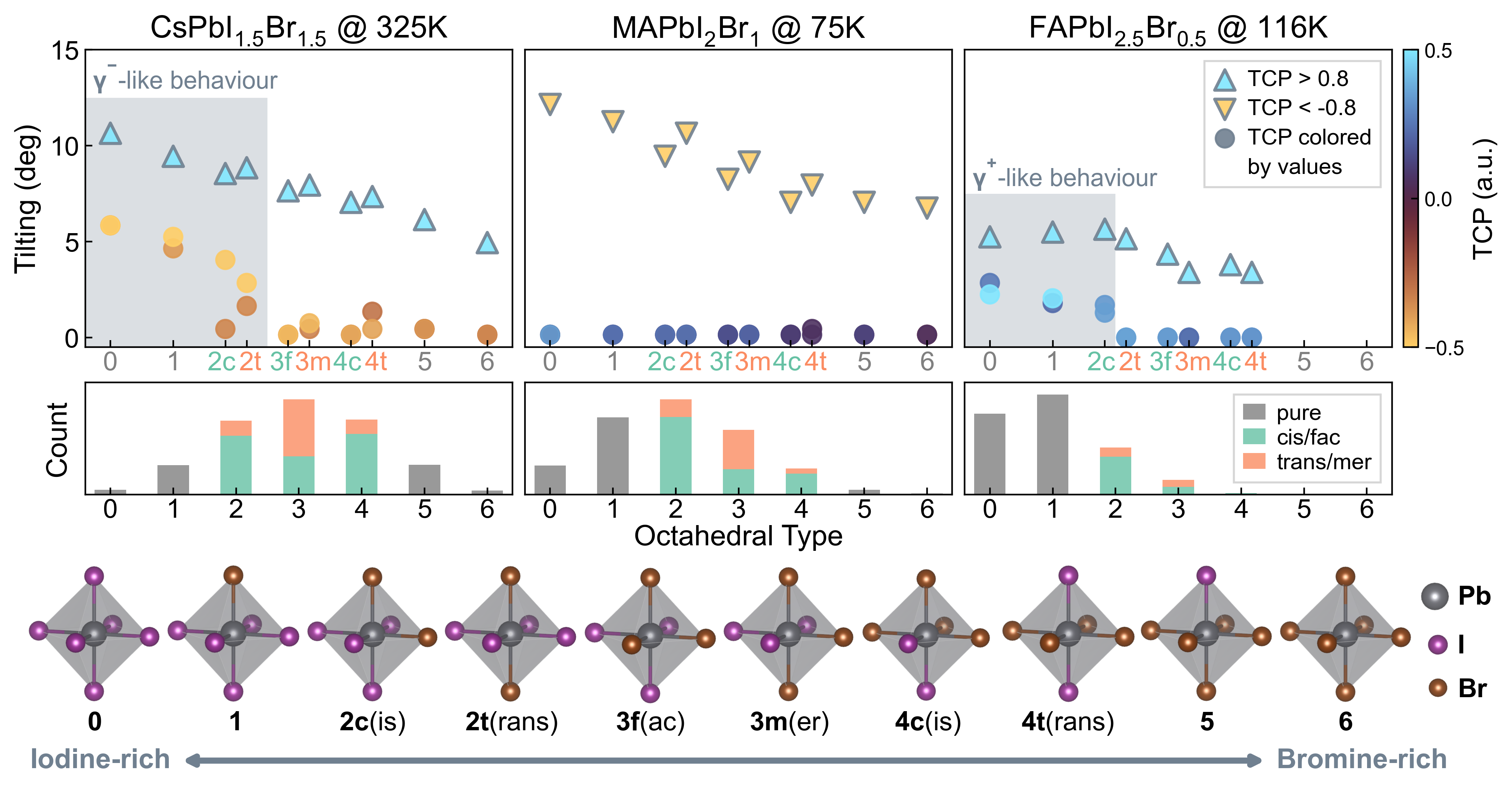}
    \caption{Octahedral tilt angles and corresponding TCP values of the three principal axes, with separated contributions from octahedra categorized into different halide configurations as in the bottom row. The relative amount of each category is also shown. The involved structures are: \ce{CsPbI_{1.5}Br_{1.5}} at 325~K, \ce{MAPbI_{2}Br_{1}} at 75~K, and \ce{FAPbI_{1.5}Br_{1.5}} at 116~K. All three structures exhibit bulk $\beta$-phase behaviour. }
    \label{res:local}
\end{figure*}

\subsection{Homogeneous Mixing}
\textbf{Local Structures in Mixed Phases.}
We now turn to partially mixed systems in their $\beta$ phases close to the $\beta$-to-$\gamma$ transition. Fig. \ref{res:local} shows three representative cases from each A-site: \ce{CsPbI_{1.5}Br_{1.5}}, \ce{MAPbI_{2}Br_{1}}, and \ce{FAPbI_{2.5}Br_{0.5}}. We classify each \ce{PbX6} octahedron according to how many Br atoms it contains (from 0 to 6), which yields ten possible local configurations. The existence of symmetry-breaking local domains caused by octahedral tilting has been commonly reported, which facilitates the formation of a temporal lower temperature phase in the bulk structure.~\cite{MA_local_order_2023_toney,dynamic_domian_milos} 
For configurations with 2, 3, or 4 bromine atoms, each case includes two isomeric sub-configurations. This classification allows for distinguishing the dynamic property contributions from different halide configurations or concentrations, offering a localized perspective on the structural behaviour. In \ce{CsPbI_{1.5}Br_{1.5}} and \ce{FAPbI_{2.5}Br_{0.5}}, I-rich octahedra exhibit more pronounced tilting akin to $\gamma$-phase, as type $0$, $1$, $2c$ and $2t$ octahedra display three non-zero tilting axes, their correlation modes also resemble their typical $\gamma$-phase. Meanwhile, Br-rich octahedra behave closer to the host $\beta$-phase. Conversely in \ce{MAPbI2Br1}, the $\gamma$-like octahedra are hardly observed near this boundary, indicating that a direct crystallization of its $\gamma$-phase from a host $\beta$-phase structure is unfavourable. This approach reveals local tilt patterns that would remain hidden if we only considered bulk-averaged properties. We also observe a clear change in the distribution of octahedral types with the global halide ratio, which approximately follows a (skew) normal distribution. For octahedra containing 2, 3 or 4 bromine atoms, the two isomeric types are not equally populated, instead, the $cis$ and $mer$ types are more dominant than the opposite. 

\begin{figure*}[htb]
    \centering
    \includegraphics[width=0.45\textwidth]{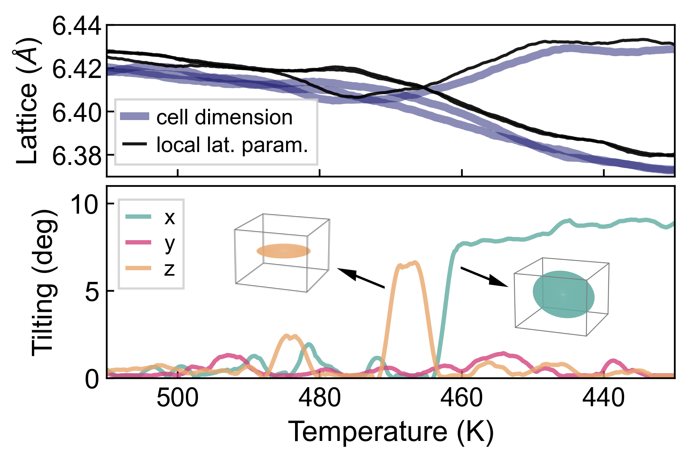}
    \caption{Cooling of \ce{CsPbI3} from its $\alpha$-phase to $\beta$-phase, observed by changes in cell dimension, local lattice parameters (relative distance between neighbouring Pb atoms~\cite{pdyna2023liang}) and tilt angles. Tilt angles reveal intermittent $\beta$-like domains that form and disappear before full phase transformation.}
    \label{res:cs_competing}
\end{figure*}

\vspace{0.6em}
\noindent
\textbf{Transient Simulations.}
Equilibrium simulations clarify stable phases, but non-equilibrium (e.g., heating or cooling) runs offer more insight into phase transformation dynamics. Fig. \ref{res:cs_competing} tracks the lattice parameters and octahedral tilt angles for a cooling run of \ce{CsPbI3} from $\alpha$- to $\beta$-phase. Both observables capture the phase change, but tilting reveals short-lived and finite-size tilt domains not visible in the lattice parameters alone. These tilt domains emerge and disappear before the stabilization of a genuine $\beta$-phase.

\subsection{Phase Diagram} %
\begin{figure*}[htb] 
    \centering
    \includegraphics[width=0.8\textwidth]{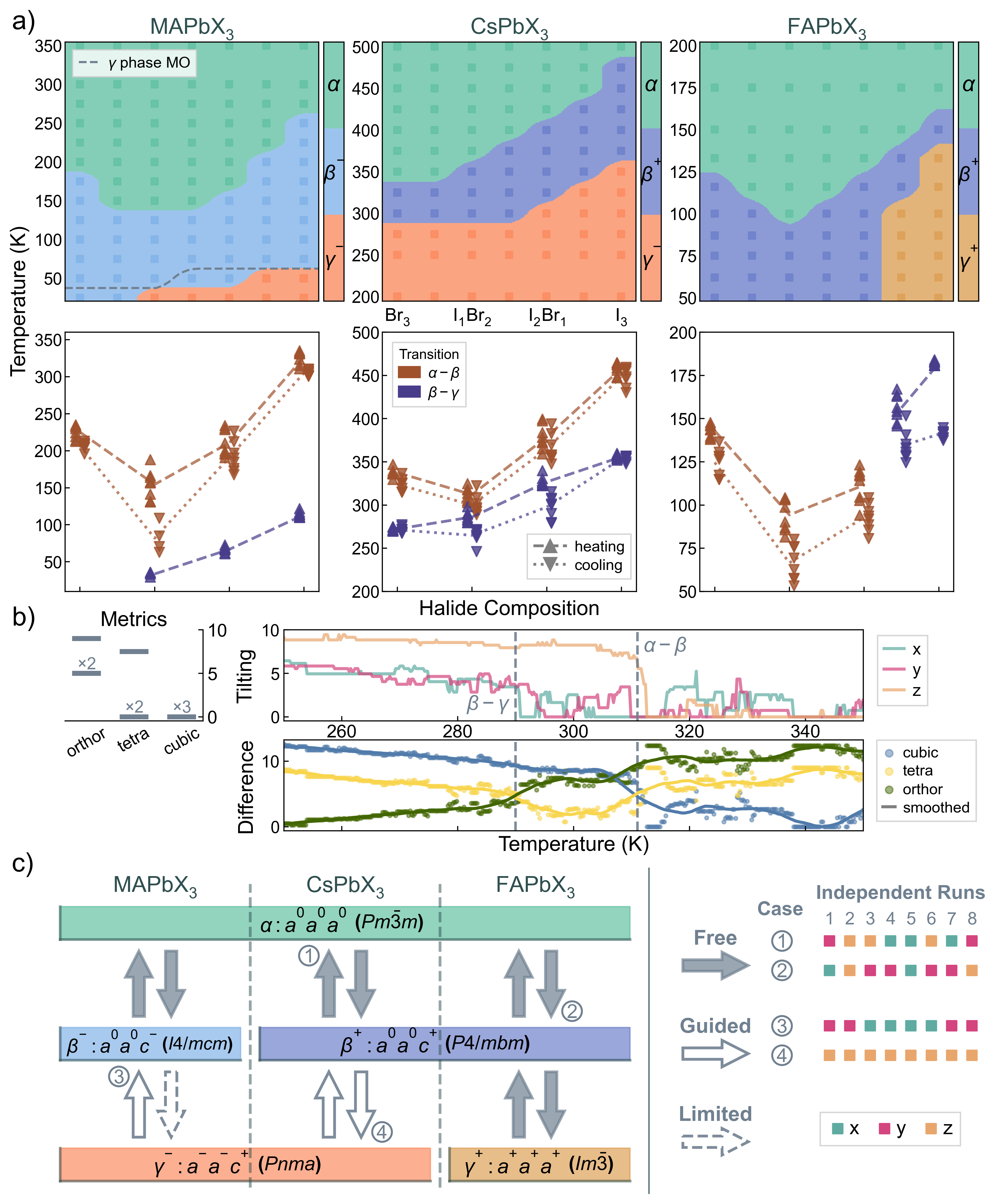}
    \caption{(a) Top row: equilibrium-based phase diagrams for \ce{CsPbX3}, \ce{MAPbX3}, and \ce{FAPbX3}. Each square is an equilibrated structure with a distinct temperature and composition. Bottom row: transition temperatures extracted from heating (upper triangles) and cooling (lower triangles). The average transition temperatures are shown with lines. (b) Example heating run from 250~K to 350~K for \ce{CsPbI1Br2}, showing a clear transition from $\gamma$- to $\beta$-, then to $\alpha$-phase. The mapping angles are shown in the top-left panel. (c) Schematic of the main transition pathways and whether such transitions are free, guided, or limited. The right panel shows the resulting orientation of the crystal within four selected transition events.}
    \label{res:phases}
\end{figure*}

Analysis of the constant-temperature MD data was used to construct phase diagrams for \ce{CsPbX3}, \ce{MAPbX3}, and \ce{FAPbX3} (top row of Fig. \ref{res:phases}a). Each dot represents a composition--temperature state, from which tilt angles and tilt correlations classify the stable phase. If a principal axis exhibits an insignificant tilt angle or a TCP value near zero (indicating neither positive nor negative correlation), it is designated as a zero-tilting axis. The number of non-zero tilting axes is then counted for phase identification. In \ce{CsPbX3} and \ce{MAPbX3}, the transition boundaries shift smoothly with halide fraction and have a reasonable agreement with experimental measurements~\cite{cspbx3_phase_diagram_nasstrom,mapbx3_phase_diagram_lehmann}. While in \ce{FAPbX3} the low-temperature $a^{+}a^{+}a^{+}$ phase is eliminated by moderate Br content, implying that such a phase may be sensitive to global softening and halide compositions. 

To probe phase transitions more precisely, we perform heating and cooling simulations at different compositions and monitor the evolution of the structural descriptors with temperature. Fig. \ref{res:phases}b shows an example heating run of \ce{CsPbI1Br2}, where the structure turned from its $\gamma$-phase to $\beta$-phase and subsequently to the high-symmetry $\alpha$-phase. The clear boundaries between these phases are obtained by comparing tilt angles against reference angles (obtained from the equilibration calculations). The phase with the highest likelihood at a temperature can be identified with the resulting angle deviation: 
\begin{equation}
\label{eq:metric}
 \Delta \theta^{j} = \sqrt{\sum_{i} \left( \theta_{i} - \hat{\theta^{j}_{i}} \right)^{2}} \quad i \in [x,y,z] \, , \, j \in [\alpha,\beta,\gamma]
\end{equation}
where $\theta_{i}$ is the tilt angle along axis $i$, and $\hat{\theta}_{i}^j$ the reference tilt for phase $j$. The phase with the smallest $\Delta \theta^{j}$ at each temperature is the most likely state. Each transition temperature is set at the intersection of $\Delta \theta$ curves.

To ensure robust sampling, eight independent MD simulations are performed to capture each of the $\beta$-to-$\gamma$ and $\alpha$-to-$\beta$ transitions during heating and cooling. These predictions, as shown in the bottom row of Fig. \ref{res:phases}a, are corroborated by transition temperatures obtained from equilibration-derived phase diagrams. Repeated simulations confirm hysteresis in halide-mixed perovskites, with heating transitions consistently occurring at higher temperatures than their cooling counterparts. This effect is more profound in the mixed compositions than in the pure endpoints. Nonetheless, the transition temperatures in the mixed halide structures manifest a wider spread, meaning that the halide mixing imposes uncertainty in the transition process, as the crystallization can be heterogeneous. Also, we note that the $\beta$-phase is absent in the transient simulations in the I-rich \ce{FAPbX3} system because it is indistinguishable with respect to the adjacent phases. 

We consider the symmetry pathways between phases by tracking how the orientation of the principal tilt axes changes during transitions. The orientation is represented by the special crystal principal axis relative to the crystal symmetry, which is the $z$-axis in the $\beta^{-}$, $\beta^{+}$ and $\gamma^{-}$ phase, and an arbitrary axis in the $\alpha$ and $\gamma^{+}$ phase. By evaluating how each transition gives rise to the orientation change, we can categorise them into three classes: \emph{free} transitions, where the orientation is not restricted to a certain pattern and is generated randomly, corresponding to the cases 1 and 2 shown in Fig. \ref{res:phases}c; \emph{guided} transitions, only a set of orientations are allowed, such as the case 3 where one of the negatively correlated axes (the two $a^{-}$ signs) become $c^{-}$ in the $\beta^{-}$ phase. Or similarly in case 4, where the $c^{+}$ axes in the $\beta^{+}$ and $\gamma^{-}$ phase have a one-to-one correspondence; as well as eventually the \emph{limited} (or effectively first-order) transitions, in which the low-temperature phase demands tilt correlations not easily introduced from the high-temperature lattice. 

\begin{figure*}[htb]
    \centering
    \refstepcounter{SIfigure}
    \includegraphics[width=0.5\textwidth]{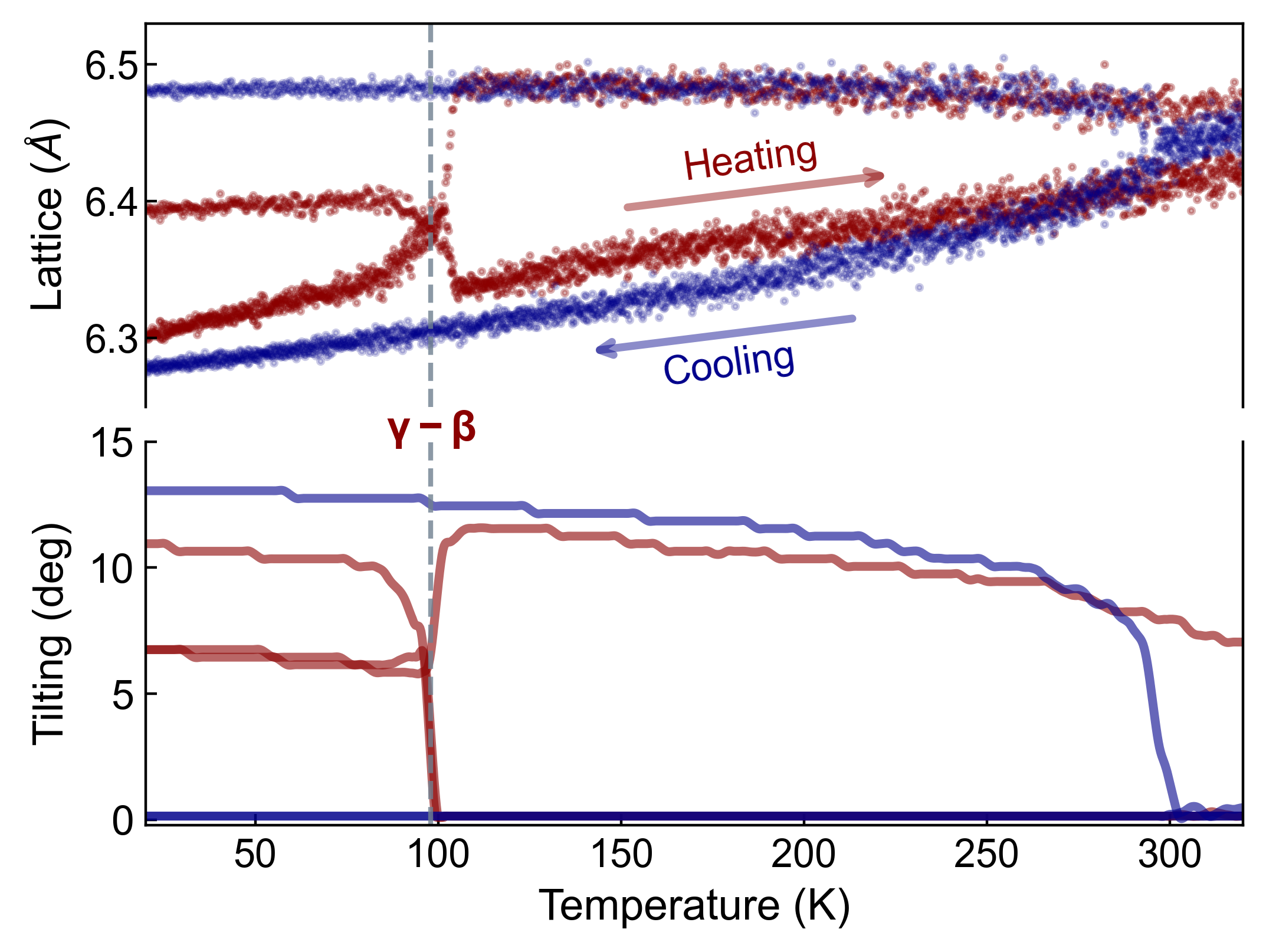}
    \caption*{Fig. \theSIfigure: Lattice parameters and octahedral tilt angles of \ce{MAPbI3} in heating and cooling MD simulation. } 
    \label{si:MA}
\end{figure*}

The symmetry-lowering process must follow a specific pathway to enable the crystallization of the low-symmetry phase from the precursor high-symmetry phase. This pathway involves introducing additional tilting correlation polarity in axes that were previously balanced ($TCP=0$). Thus among the twelve phase transition processes, this condition is satisfied in all cases except for the $\beta$-to-$\gamma$ transition in \ce{MAPbX3} (the low-temperature heating and cooling runs for \ce{MAPbX3} are shown in Fig. \ref{si:MA}). This aligns perfectly with our previous finding in Fig. \ref{res:local} which shows minimal $\gamma$-like domains in \ce{MAPbX3}. This observation agrees well with previously established works based on symmetry group relationships~\cite{BaZrS3_prakriti,howard1998group}, while providing an intuitive interpretation to the (first- and second-order) nature of transitions and their connection to structural evolution. Such distinct low-temperature behaviours arising from the various A-sites will then be the focus of the following discussion. 

\begin{figure*}[htb]
    \centering
    \includegraphics[width=1.0\textwidth]{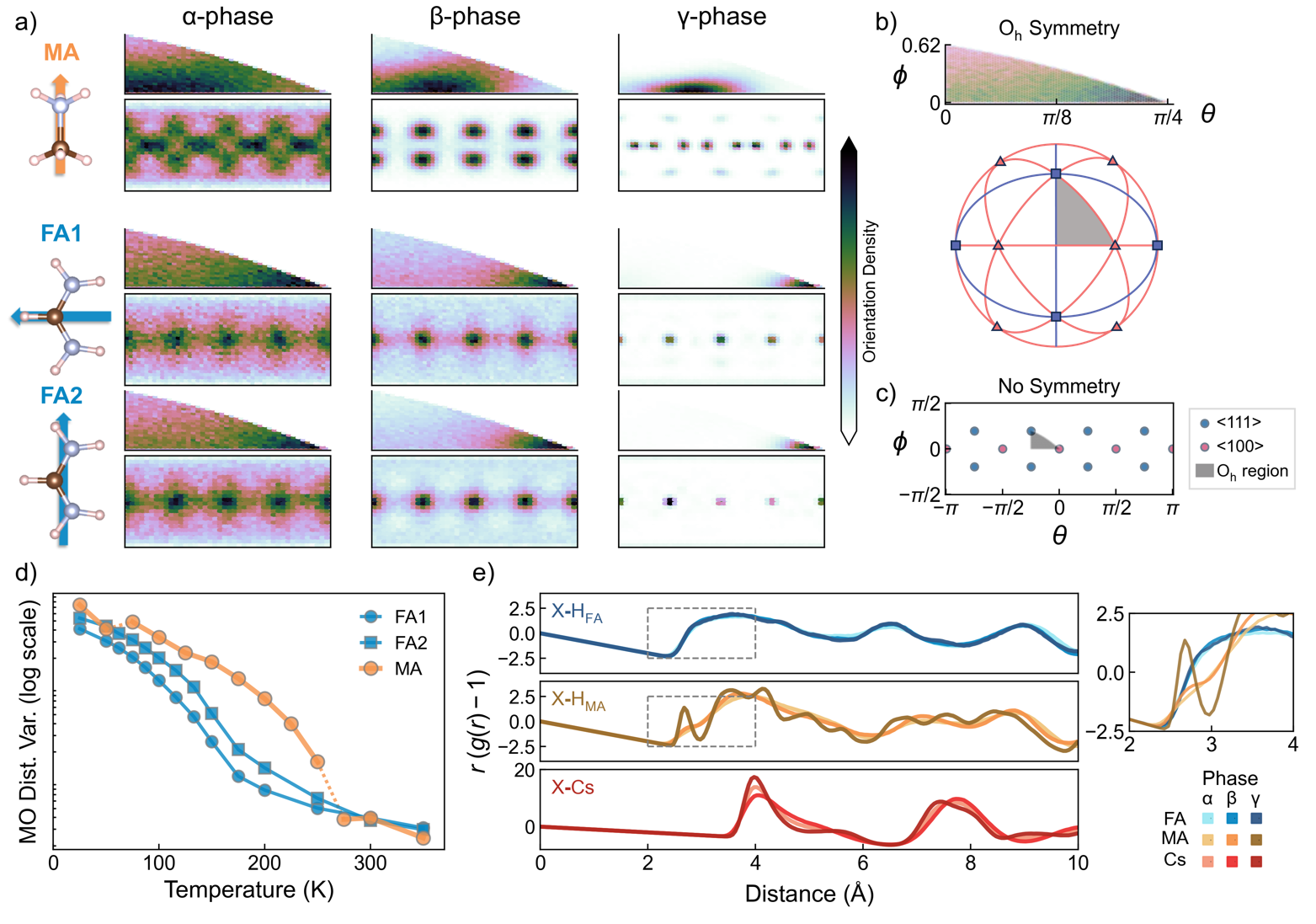}
    \caption{(a) Molecular orientation (MO) distribution of \ce{MA+} at 50~K ($\gamma$), 150~K ($\beta$), and 300~K ($\alpha$), and \ce{FA+} at 100~K ($\gamma$), 150~K ($\beta$), and 200~K ($\alpha$), orientation of FA molecule is described with two distinct orientation vectors. The orientation vectors are visualized in two spherical coordinate projections: one accounting for full $O_h$ symmetry (top sub-panels) and one without symmetry (bottom sub-panels). A 3D visualization is shown in Fig. \ref{si:MO_3D}. (b) Illustration of $O_h$-equivalent orientations in 3D space.~\cite{hhp_molecular_ferroelectric_frost} (c) No-symmetry projection scheme highlighting $\langle 111 \rangle$ and $\langle 100 \rangle$ directions. Molecular orientations of the A-site molecule are projected onto horizontal (azimuthal angle $\theta$) and vertical (polar angle $\phi$) axes. (d) Variance of molecular orientation distribution as a function of temperature. (e) Pair distribution functions for hydrogen--halide (\ce{X-H_{FA}}, \ce{X-H_{MA}}) and \ce{X-Cs} pairs across phases, with a zoomed-in region.}
    \label{res:mo}
\end{figure*}

\vspace{0.6em}
\noindent
\textbf{A-Site Effects on Phase Behaviour.}
The dynamic behaviour of perovskites is influenced by the choice of the A-site cation. The caesium cation, due to its small ionic radius, exhibits the weakest interaction with the surrounding octahedral framework. Consequently, the cubic phase of the \ce{CsPbX3} system remains stable only at very high temperatures, as tilting readily occurs in response to the large cavity size. In contrast, both organic A-site cations have larger ionic radii than Cs, leading to lower transition temperatures to low-symmetry phases. 

Fig. \ref{res:mo}a compares \ce{MA+} and \ce{FA+} orientations in \ce{MAPbI3} and \ce{FAPbI3} across the $\alpha$, $\beta$, and $\gamma$ phases. For MA, despite considerable similarities in orientation within the $O_{h}$ space across the three phases, the full 3D $\gamma$-phase orientation distribution differs substantially from the higher-temperature phases, which explains why it fails to nucleate under typical cooling protocols with accessible cell sizes and cooling rates, and why the coexistence of both phases in the same structure is impossible (Fig. \ref{res:local}). Because a strong rearrangement of molecules is compulsory in this case, such a finding is highly consistent with other DFT calculations on kinetics~\cite{mapi_beta_gamma_kinetics_wu}, neutron scattering experiment~\cite{ma_dynamics_leguy} and MD simulations~\cite{picosecond_lifetime_MA_Eduardo,vaspmlff_other_mapb_halides}. The $\beta$-phase orientation is a subset of the $\alpha$-phase, so the transition is much easier. By contrast, FA molecules largely retain the same preferred $\langle 100 \rangle$ orientation direction with only diminishing thermal fluctuations at lower temperatures, with good agreement with DFT calculations~\cite{dynamics_molecule_iodide_fabini}. This continuity facilitates phase changes that do not rely on a large rearrangement of the A-site orientation. Both the discontinuity in distribution variance (calculated with Equation~\ref{eq:mo_var} in the SI) and the pair distribution functions (PDFs) support these observations: the $\gamma$-phase in MA-based perovskites shows distinct hydrogen--halide interactions, while FA-based perovskites maintain similar PDFs across phases (Fig. \ref{res:mo}d,e). This explains why the softening effect is most profound in the FA-containing systems, as the FA cation imposes a much weaker influence on the tilting and phase transition of the \ce{BX6} octahedral framework. Thus, such systems are prone to the softening effect where the phases are mostly driven by the octahedral tilting.

\begin{figure*}[htb]
    \centering
    \includegraphics[width=0.85\textwidth]{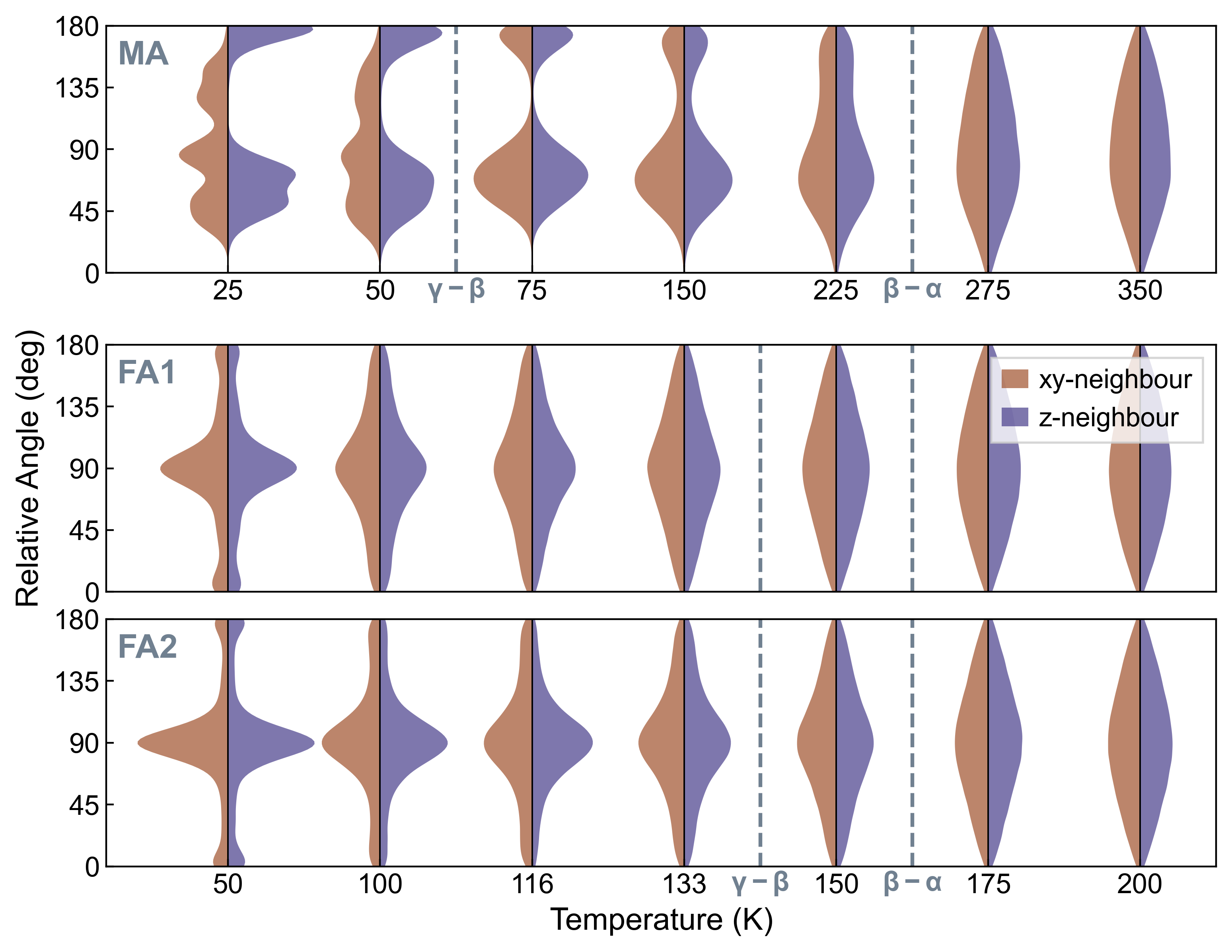}
    \caption{Distribution of relative angles between molecular vectors of MA, FA1 and FA2 vectors and their corresponding neighbours, in the $xy$-plane (brown) and along the $z$-direction (dark purple) within the crystal srtucture. Data is collected for various temperatures in \ce{MAPbI3} and \ce{FAPbI3}. The relevant phase transition temperatures are indicated with vertical dashed lines. }
    \label{res:mo_order}
\end{figure*}

Such a contrast between MA and FA molecules also becomes evident in their three-dimensional alignments. Figure~\ref{res:mo_order} illustrates the relative angles of each molecular vector with its nearest neighbour, accounting for crystal symmetry and directions. In FA-based systems, both the FA1 and FA2 vectors strongly prefer to point along the $\langle 100 \rangle$ directions, indicating that their relative angles should theoretically be 0, 90, or 180 degrees. Indeed, the observed data confirm a pronounced preference for a 90-degree relative angle, growing more dominant at lower temperatures while remaining isotropic across all directions. Notably, this behaviour holds regardless of the crystal phase, implying minimal molecular rearrangement barriers during phase transitions. 

By contrast, MA-based systems display markedly different molecular alignment schemes among their various phases. Although the $\alpha$-phase and $\beta$-phase exhibit relatively similar modes of alignment, the $\gamma$-phase adopts drastically distinct orientations relative to higher-temperature phases, most notably featuring an additional rearrangement near 135 degrees in the $xy$-plane and moderate shifts in the 45-to-90-degree range. Coupled with slower rotational dynamics at lower temperatures, these changes significantly impede the $\beta$-to-$\gamma$ transition in MA systems, in agreement with the static observations shown in Fig. \ref{res:mo}. This also suggests that our previous determination of the \ce{MAPbX3} phase diagram shown in Fig. \ref{res:phases} is incomplete, as the initial structure contains randomly rotated MA molecules, whereas the phase transition is strongly related to the molecular motion. When the molecules in the initial structures are aligned as in the pattern found in equilibrated $\gamma$ phase, the $\gamma$ phase can be found in a wider window of temperatures and halide compositions (see dashed line in top left panel of Fig. \ref{res:phases}).  

\begin{figure*}[htb]
    \centering
    \includegraphics[width=1.0\textwidth]{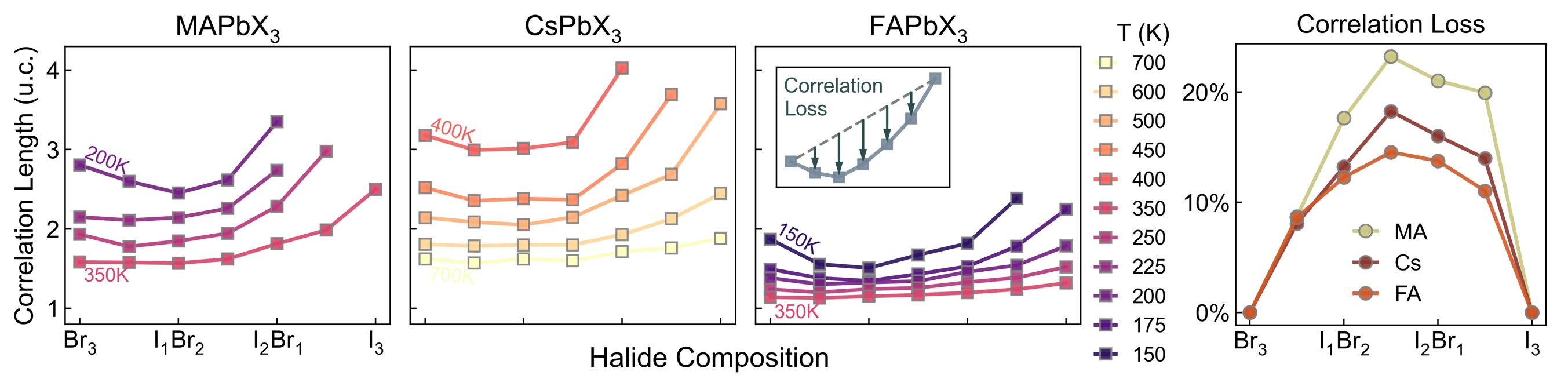}
    \caption{Correlation lengths for octahedral tilting in the cubic ($\alpha$) phase of \ce{MAPbX3}, \ce{CsPbX3}, and \ce{FAPbX3} at different compositions and temperatures. Right panel: percentage reduction in correlation length due to halide mixing.}
    \label{res:corr_lengths}
\end{figure*}

\vspace{0.6em}
\noindent
\textbf{Tilting Correlation Lengths.}
One way to interpret these phase boundaries is through correlation length. The formation of the $\beta$-phase from the bulk untilted $\alpha$-phase can be conceptualized as the establishment of an infinitely large tilt domain along one direction. By fitting the spatial decay of tilt angles to an exponential, we extract correlation lengths for each composition in the cubic ($\alpha$) phase (Fig. \ref{res:corr_lengths}). At the same temperature, \ce{CsPbX3} shows longer correlation lengths than organic-cation systems because the larger cavity around \ce{Cs+} more readily supports octahedral tilting. Meanwhile, \ce{FAPbX3} has the shortest correlation lengths, a trend that correlates inversely with the size of the A-site cation. The pure iodine endpoint possesses a longer correlation length than bromine because the longer Pb-I bond promotes stronger octahedral tilting. Halide mixing further disrupts tilting continuity: the correlation length for mixed compositions is consistently below the linear interpolation between the pure endpoints, likely because local fluctuations in halide composition impede coherent tilt propagation. Therefore, especially within the organic compounds, the $\alpha$-to-$\beta$ phase transition temperatures in mixed halide systems are lower compared to those of the pure endpoints, consistent with the shorter correlation lengths. 

\vspace{0.6em}
\noindent
\textbf{Thermodynamic Stability.}

While the thermodynamics of mixing for crystals is usually estimated from static calculations, we can employ the ensemble energies from our MD simulations for this purpose.
By performing consistent simulations of the mixed and end-member compounds, we can obtain a direct estimate of the enthalpy of mixing $\Delta H_{mix}$. 

Assuming an ideal entropy of mixing $\Delta S_{mix}$ based on the chemical composition and available sites, we define the free energy of mixing $\Delta G_{mix}$:
\begin{equation}
\label{eq:mixing}
\begin{split}
 \Delta S_{mix} &= -k_{B} \left( x_A \ln x_A + x_B \ln x_B \right) \\
 \Delta G_{mix} &= \Delta H_{mix} - T\Delta S_{mix}
\end{split}
\end{equation}
Here, $k_{B}$ is the Boltzmann constant expressed in $eV / K$, and $x_A$ and $x_B$ represent the occupation fractions of the binary mixing sites, in this context, iodine and bromine. 

\begin{figure*}[htb]
    \centering
    \includegraphics[width=0.8\textwidth]{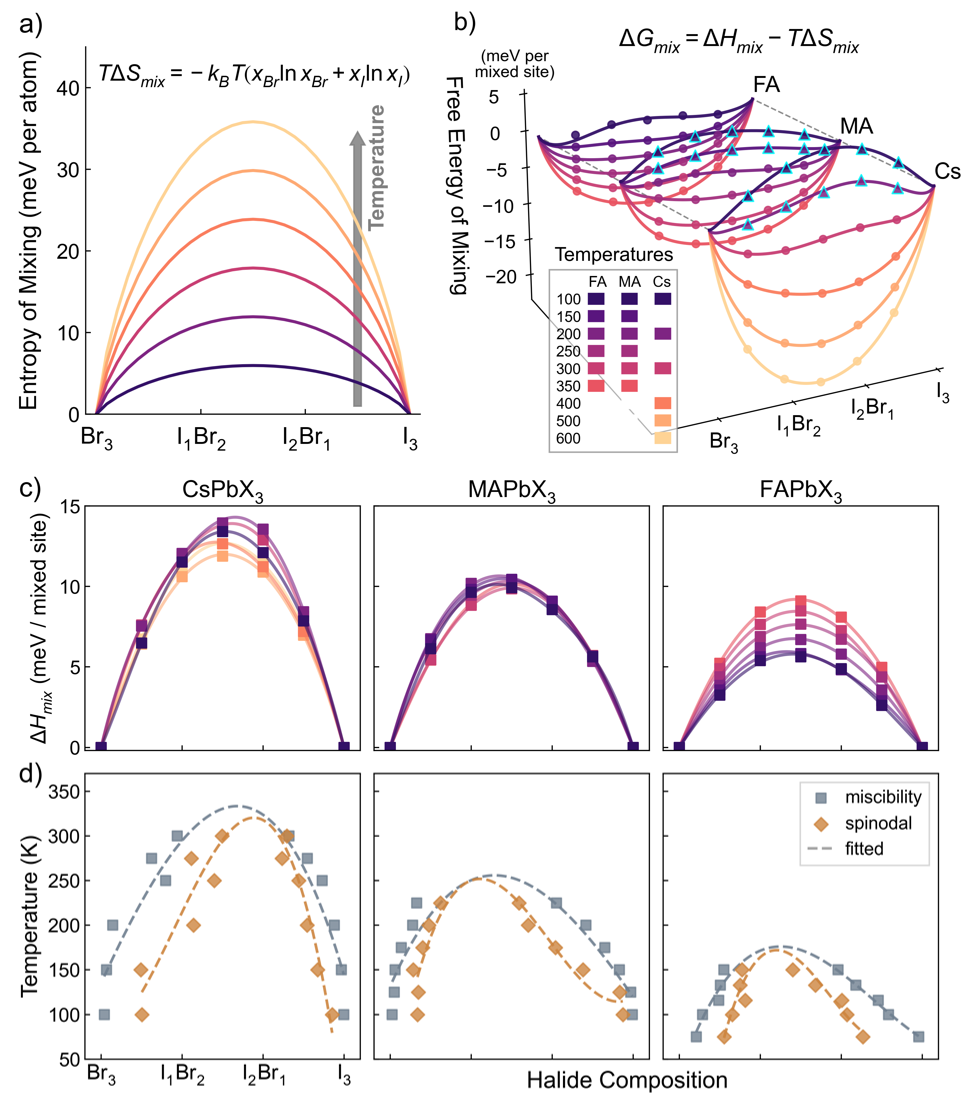}
    \caption{(a) Entropy of mixing calculated from Equation~\ref{eq:mixing}. (b) Free energy of mixing for each system. The scatter points denote values predicted by force field, where values above zero are indicated by upper triangles. The lines are plotted from fitted functions. (c) Enthalpy of mixing, the squares are the raw data and the lines represent the fitted values. (d) Phase diagram of miscibility gap and spinodal decomposition, obtained from numerical fitting to common tangents, indicating regions of metastable and unstable mixing of halides. }
    \label{res:energies}
\end{figure*}

Fig. \ref{res:energies}b shows $\Delta G_{mix}$ versus composition for each system. At low temperatures, \ce{CsPbX3} and \ce{MAPbX3} have slightly positive $\Delta G_{mix}$, suggesting an intrinsic driving force for halide segregation. By contrast, \ce{FAPbX3} remains negative over the temperature range studied, implying a stronger tendency for homogeneous mixing. Furthermore, by fitting fifth-order polynomial functions to the predicted enthalpy of mixing (containing seven data points with varying halide composition), one can also estimate the first and second derivative of the free energy of mixing, allowing the prediction of miscibility gap and spinodal decomposition as shown in Fig. \ref{res:energies}c, in good agreement to well-established DFT predictions~\cite{all_asite_segregation_chen,MA_segregation_choe}. Consequently, the FA system possesses the highest stability among the three solid solutions. 

\subsection{Heterogeneous Mixing}
\begin{figure*}[htb]
    \centering
    \includegraphics[width=0.99\textwidth]{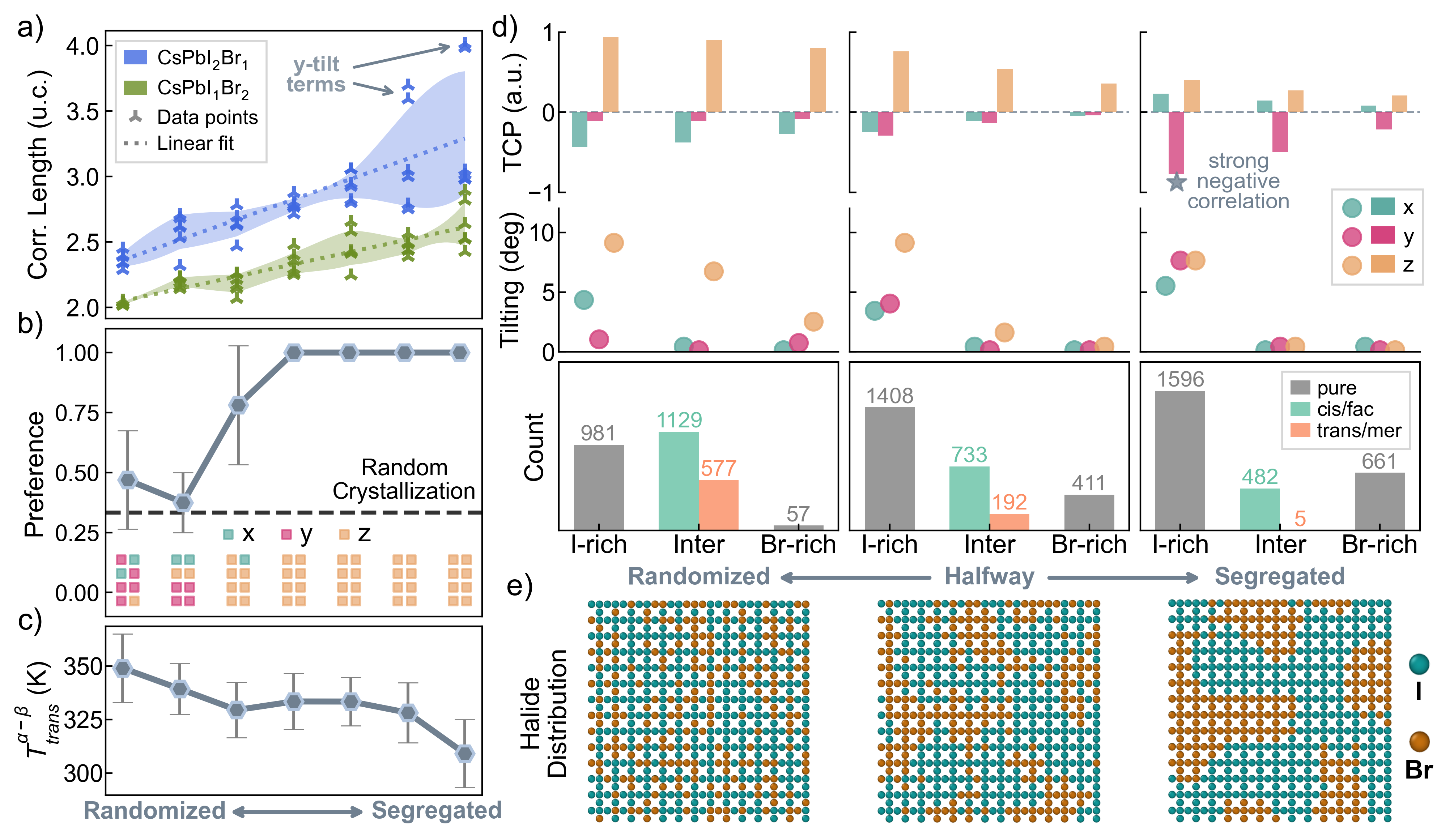}
    \caption{The effect of the degree of halide segregation on various properties. (a) Correlation length of tilting in parallel directions for \ce{CsPbI2Br1} and \ce{CsPbI1Br2} at 500 K. (b) Crystallization direction preference of \ce{CsPbI2Br1}. The crystallization axes from eight independent runs are shown in the lower part of the plot. (c) $\alpha$-to-$\beta$ phase transition temperature of \ce{CsPbI2Br1}. (d) TCP values, tilting angles, and counts of octahedral types for as-randomised, half-segregated, and segregated \ce{CsPbI2Br1} structures at 350~K, at which the host structure is in the $\beta$-phase. (e) Schematics of halide distributions. }
    \label{res:segregate}
\end{figure*}

Despite halide segregation and diffusion not being our primary focus, the direct simulation of locally clustered halides is an important case as local ordering is often observed in real systems, especially under illumination. A custom Monte Carlo algorithm promotes I--Br pair swaps that increase local halide concentration, creating distinct I-rich and Br-rich domains. Intermediate structures with partial segregation further bridge random and strongly segregated distributions (Fig. \ref{res:segregate}e). 

Fig. \ref{res:segregate}a shows that tilting correlation lengths grow as halides become more locally clustered. All nine components of the correlation length tensor (three normal terms $\xi_{\alpha,\beta}$ when $\alpha=\beta$, and six parallel terms when $\alpha \ne \beta$, explained in Eq.~\ref{eq:corr_length2}) increase approximately linearly with the degree of segregation. Near-complete segregation at 500~K in \ce{CsPbI2Br1} yields symmetry-breaking behaviour where the two correlation length components related to the $y$-tilt become significantly dominant. This suggests that, as halides segregate, crystallization into low-symmetry phases is restricted to specific orientations. To verify this hypothesis, repetitive transient simulations were performed on the segregated structures to analyze the orientations in which the $\beta$-phase forms. The crystallization preference $G$ is quantified using the following metric:
\[
    G = \frac{1}{n} \sum^{n}_{i} \sum^{n}_{j} g_{ij} \: , \quad \text{where} \: \:  g_{ij}= 
\begin{dcases}
    \frac{1}{n},    & \text{if } k_{i} = k_{j}\\
    0,              & \text{otherwise}
\end{dcases}
\]
Here, $i$ and $j$ denote the indices of transient simulation samples, and $k_{i}$ represents the crystallization orientation in sample $i$. The value of $G$ ranges from $\frac{1}{n}$, indicating random crystallization, to 1, indicating fully controlled crystallization. As segregation intensifies, the computed $G$ approaches unity, meaning the phase transition becomes strongly predetermined by the structure. At intermediate segregation levels, the local structures promote tilting correlations along a specific axis, leading to directional crystallization during cooling (Fig. \ref{res:segregate}b). This finding implies that in real-world scenarios, such as light-induced halide segregation, similar symmetry-breaking effects could complicate device performance.

Furthermore, Fig. \ref{res:segregate}c indicates that halide segregation lowers the phase transition temperature to lower-symmetry phases. To investigate this phenomenon, we examine the behaviour of domains enriched in either bromine or iodine. Fig. \ref{res:segregate}d presents the tilting angles and tilt correlation polarity (TCP) values for I-rich (octahedral types $0$ and $1$), Br-rich (types $5$ and $6$), and intermediate regions (covering the remaining types), along with their respective populations. As segregation intensifies, the fraction of strongly I-rich and Br-rich octahedra increases, while the $cis$ and $fac$ isomeric types dominate at the boundaries between these regions. Both trends agree with geometric intuition.

In the randomized (homogeneous) structure, the overall tilting is consistent with a $\beta^{+}$ phase. However, as segregation proceeds, the I-rich domains develop strong tilts, whereas Br-rich areas adopt an $\alpha$-like, low-tilt configuration. When segregation is nearly complete, an atypical $b^{-}$ tilting mode arises in the I-rich regions with disordered tilts in the other two directions, driven by the constraints imposed by the surrounding low-tilt domains. This phenomenon of distinct structural behaviours in segregated I-rich and Br-rich regions is also captured through X-ray diffraction measurements.~\cite{exp_trap1,segregation_xray_halford} We note that the crystal orientation and correlation mode remain sensitive to the initial conditions and Monte Carlo steps, yet the fundamental emergence of domain boundaries, and their impact on tilting, should be prevalent across different segregation scenarios. Consequently, the local formation of incompatible tilting modes in I-rich areas, coupled with disruptions to tilt propagation at domain interfaces, explains the reduction in transition temperature under higher segregation.


\section{Conclusion}

The interplay among the A-site cation, halide composition, and local atomic environments critically shapes phase stability and transitions in mixed halide perovskites. By combining on-the-fly DFT data selection, machine learning potentials and a perovskite structural analysis framework, we identified the stable phases and constructed phase diagrams of \ce{CsPbX3}, \ce{MAPbX3}, and \ce{FAPbX3}. Among the three A-site cations studied, \ce{MA+} leads to a unique symmetry where the $\beta$-to-$\gamma$ transition in \ce{MAPbX3} is effectively “forbidden” in a finite cell and time, as it requires symmetry-breaking tilts and a significant rearrangement of the MA cations that cannot be accommodated by the $\beta$-phase. In addition, we show how small compositional variations and halide arrangements can significantly alter octahedral tilting, correlation lengths, and phase transformation pathways. While halide mixing weakens the overall tilting correlation by introducing compositional heterogeneity, local segregation fosters a strong global tilting correlation that can produce anomalous tilting modes. However, these segregated boundaries ultimately impede coherent phase transitions. Overall, these findings highlight the value of high-accuracy simulations for capturing multi-scale structural responses and guide composition–structure tuning to engineer stable, high-performance perovskite materials.

\newpage

\section{Code Availability}
The \textsc{PDynA} package used in this work is open-source and available online at \url{https://github.com/WMD-group/PDynA} (DOI: 10.5281/zenodo.7948045). A repository with full training data and force field paramaters will also be provided. 

\section{Acknowledgements}
We thank Z. Zeng, M. Dubajic, and S. Stranks
for their helpful discussions. J. K. acknowledges support from the Swedish Research Council (VR) program 2021-00486. AW thanks the Leverhulme Trust (RPG-2021-191) for funding. Via our membership of the UK's HEC Materials Chemistry Consortium, which is funded by EPSRC (EP/X035859/1), this work used the ARCHER2 UK National Supercomputing Service (http://www.archer2.ac.uk). We are also grateful to the UK Materials and Molecular Modelling Hub for computational resources, which is partially funded by EPSRC (EP/T022213/1, EP/W032260/1 and EP/P020194/1).

\newpage
\section{Supporting Information} 

\begin{table}[ht]
\centering
\refstepcounter{SItable}
\caption*{Table \theSItable: Root mean squared errors of energy, forces and stress tensors of the trained machine learning force fields }
\begin{tabular}[t]{c@{\hskip 0.4in}c@{\hskip 0.3in}c@{\hskip 0.3in}c}
\multirow{2}{*}{Material} & \multicolumn{3}{c}{RMSE error on test set} \\
\cline{2-4}
 & Energy (\text{meV/atom}) & Force (meV/\AA) & Stress (meV/\AA$^{3}$) \\
\hline
\hline
\textbf{\ce{CsPbX3}} & \textbf{$0.07$} & \textbf{$2.13$} & \textbf{$0.06$} \\ 
\ce{CsPbBr3} & \textbf{$0.04$} & \textbf{$1.50$} & \textbf{$0.08$}  \\
\ce{CsPbI3} & \textbf{$0.15$} & \textbf{$2.06$} & \textbf{$0.04$} \\
\textbf{\ce{MAPbX3}} & \textbf{$0.06$} & \textbf{$3.52$} & \textbf{$0.32$} \\
\ce{MAPbBr3} & \textbf{$0.81$} & \textbf{$2.70$} & \textbf{$0.27$} \\
\ce{MAPbI3} & \textbf{$0.69$} & \textbf{$2.80$} & \textbf{$0.23$} \\
\textbf{\ce{FAPbX3}} & \textbf{$0.07$} & \textbf{$4.64$} & \textbf{$0.30$} \\
\ce{FAPbBr3} & \textbf{$0.69$} & \textbf{$3.29$} & \textbf{$0.29$} \\
\ce{FAPbI3} & \textbf{$0.56$} & \textbf{$3.57$} & \textbf{$0.26$} \\
\hline
\end{tabular}
\label{errors}
\end{table}%

\newpage

\vspace{0.8em}
\noindent
\textbf{Numerical Random and Special Quasi-random Structures}

\begin{figure*}[htb]
    \centering
    \refstepcounter{SIfigure}
    \includegraphics[width=0.8\textwidth]{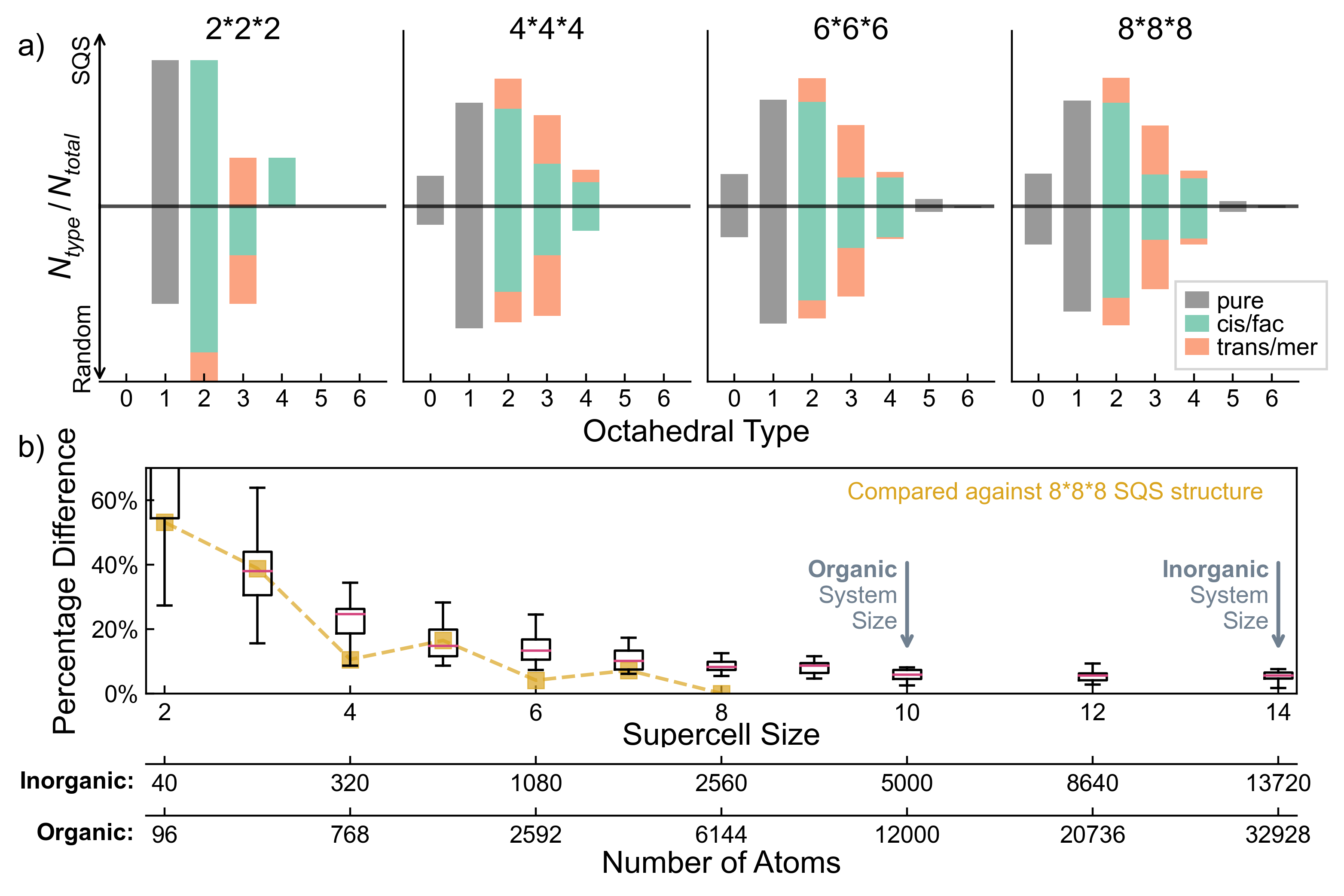}
    \caption*{Fig. \theSIfigure: (a) Fraction of octahedra in each halide configuration, generated with SQS method (upper panel) and numerical random placement (lower panel), within a \ce{CsPbI_{2.5}Br_{0.5}} structures of supercell sizes of $2\times2\times2$, $4\times4\times4$, $6\times6\times6$, and $8\times8\times8$. (b) Percentage difference of configurational occupation versus system size, compared to the $8\times8\times8$ SQS structure. The yellow line represents the difference between the smaller cell SQS structure and the $8\times8\times8$ SQS structure, and the box plot depicts the difference between 20 random structures and the $8\times8\times8$ SQS structure. }
    \label{si:system_size}
\end{figure*}

In the main text, the distribution of halides in the DFT training structures is randomised with the use of the special quasi-random structure (SQS) method, aiming for the generation of the most diverse local configurations. However, for the large-scale MD simulations, a simple numerical random placement of halides is implemented due to the high computational cost of SQS calculation at such scales. Fig. \ref{si:system_size} shows that the difference between random placement and SQS method vanishes quickly above $8\times8\times8$ supercell. 

\newpage

\vspace{0.8em}
\noindent
\textbf{Segregation Parameter}

\begin{figure*}[htb]
    \centering
    \refstepcounter{SIfigure}
    \includegraphics[width=0.8\textwidth]{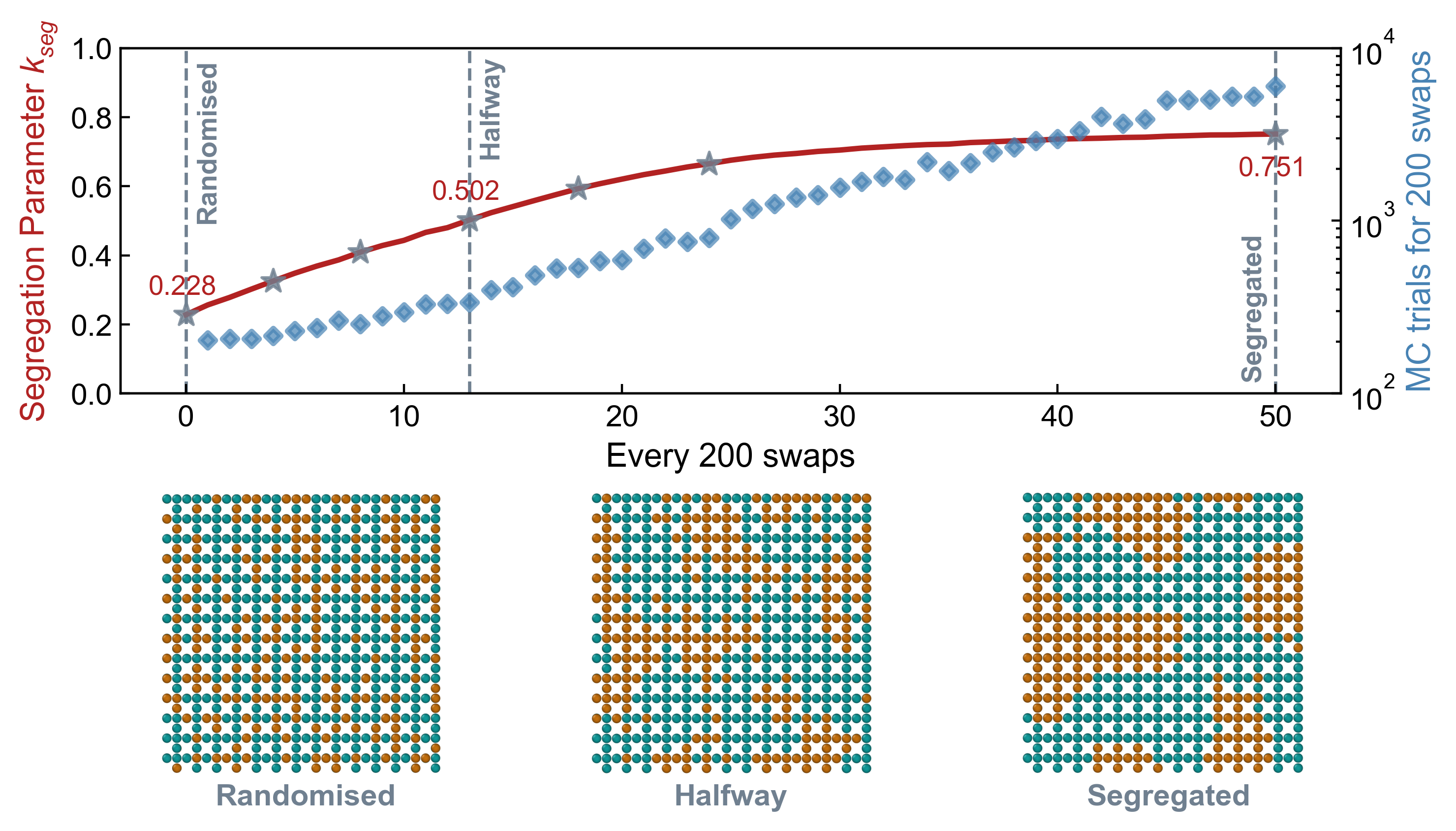}
    \caption*{Fig. \theSIfigure: Evolution of the segregation parameter $k_{seg}$ and the number of trials for 200 successful swaps during the Monte Carlo process. The 7 profiles selected in Fig. \ref{res:segregate}a--c are indicated with grey stars, and the three representative structures (\emph{Randomised}, \emph{Halfway}, and \emph{Segregated}) are outlined.}
    \label{si:segregate}
\end{figure*}

Here we show how the segregation parameter $k_{seg}$ changes with the Monte Carlo process and how it relates to the multiple segregated structures in Fig. \ref{si:segregate}. 

\newpage

\vspace{0.8em}
\noindent
\textbf{Tilt angles and TCP values of Selected Profiles}

Within the same region of a stable phase, the tilt angles and TCP values would still change with temperature. Here we show three fixed-composition profiles, one from each perovskite system. 

\newpage

\vspace{0.8em}
\noindent
\textbf{MO Distribution Variance}

The molecular orientation distribution is evaluated by binning the orientation vectors into a set of evenly distributed hexagonal bins in the three-dimensional space, which quantitatively determines the degree of concentration of orientations. Then the distribution variance $\sigma^{2}_{MO}$ can be simply defined as the variance of the normalized bin counts, 

\begin{equation}
\label{eq:mo_var}
 \sigma^{2}_{MO} = \frac{1}{N_{bin}} \sum_{\theta, \phi} \{m(\theta, \phi) - \bar{m}\}^{2}
\end{equation}

where $N_{bin}$ is the number of 3D bins, $\theta$ and $\phi$ are the azimuthal and polar angles, respectively. $m(\theta, \phi)$ represents the occupation ratio of orientation vectors located in the bin at angle $\theta$ and $\phi$, and $\bar{m}$ is the average occupation. 

\begin{figure*}[htb]
    \centering
    \refstepcounter{SIfigure}
    \includegraphics[width=0.8\textwidth]{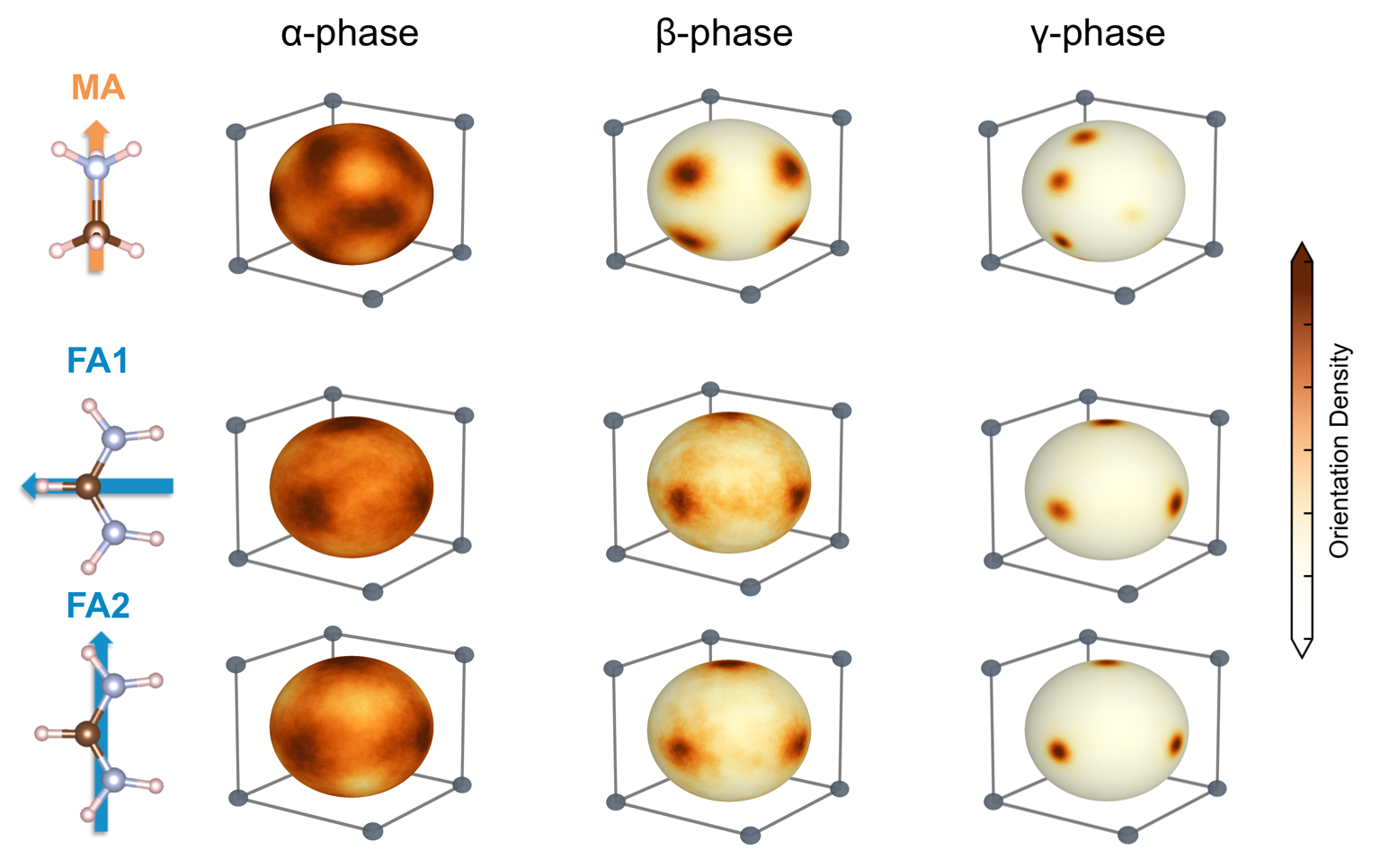}
    \caption*{Fig. \theSIfigure: Three-dimensional visualization of MO of MA and FA molecules under various phases. } 
    \label{si:MO_3D}
\end{figure*}

\newpage


\bibliography{REF.bib}

\end{document}